\date{}
\documentclass[10pt,aps,prl,twocolumn,superscriptaddress]{revtex4-1}
\usepackage{graphicx,psfrag,amsmath,amssymb,amsfonts,color}
\usepackage[T1]{fontenc}
\usepackage{natbib}
\usepackage[hidelinks]{hyperref}
\usepackage{upgreek}
\usepackage{color}
\usepackage{ulem}
\usepackage[nice]{nicefrac}
\usepackage{siunitx}

\newcommand{\beq}{\begin{equation}}
\newcommand{\eeq}{\end{equation}}
\newcommand{\bea}{\begin{eqnarray}}
\newcommand{\eea}{\end{eqnarray}}

\begin{document}

\title{Spin-selective coherent light scattering from ion crystals}

\author{Maurizio Verde}
\affiliation{QUANTUM, Institut für Physik, Universität Mainz, Staudingerweg 7, 55128 Mainz, Germany}

\author{Ansgar Schaefer}
\affiliation{QUANTUM, Institut für Physik, Universität Mainz, Staudingerweg 7, 55128 Mainz, Germany}
\altaffiliation{Corresponding author: \href{mailto:bzenz@uni-mainz.de}{bzenz@uni-mainz.de}}

\author{Benjamin Zenz}
\altaffiliation{Corresponding author: \href{mailto:bzenz@uni-mainz.de}{bzenz@uni-mainz.de}}
\affiliation{QUANTUM, Institut für Physik, Universität Mainz, Staudingerweg 7, 55128 Mainz, Germany}

\author{Zyad Shehata}
\affiliation{AG Quantum Optics and Quantum Information, Friedrich-Alexander-Universität Erlangen-Nürnberg,
Staudtstraße 1, 91058 Erlangen, Germany}

\author{Stefan Richter}
\affiliation{AG Quantum Optics and Quantum Information, Friedrich-Alexander-Universität Erlangen-Nürnberg,
Staudtstraße 1, 91058 Erlangen, Germany}

\author{Christian T. Schmiegelow}
\affiliation{Departamento de Física, FCEyN, Universidad de Buenos Aires and IFIBA, CONICET, Pabellón 1, Ciudad Universitaria, 1428 Ciudad de Buenos Aires, Argentina}

\author{Joachim von Zanthier}
\affiliation{AG Quantum Optics and Quantum Information, Friedrich-Alexander-Universität Erlangen-Nürnberg,
Staudtstraße 1, 91058 Erlangen, Germany}

\author{Ferdinand Schmidt-Kaler}
\affiliation{QUANTUM, Institut für Physik, Universität Mainz, Staudingerweg 7, 55128 Mainz, Germany}
\affiliation{Helmholtz-Institut Mainz, Staudingerweg 18, 55128 Mainz, Germany}

\begin{abstract}
We study coherent light scattering from linear crystals with up to twelve $^{40}\text{Ca}^+$ ions, acting as single photon emitters. Light-scattering is induced by two-photon laser excitation, starting from the S$_{1/2}\rightarrow$ D$_{5/2}$ quadrupole transition at 729~nm followed by the D$_{5/2}\rightarrow$ P$_{3/2}$ dipole transition at 854~nm, from where the ions decay back to the S$_{1/2}$ ground state via emission of a photon near 393~nm. We realize spin-selective excitation from the Zeeman-split ground states S$_{1/2}$, m$= \pm 1/2$, of the $\text{Ca}^+$ ions and observe in the far field spin-dependent interference patterns displaying the spin textures of the ion crystals. We investigate their dynamics by measuring the temporal evolution of the spatial Fourier frequencies of the observed patterns.

\end{abstract} 

\maketitle

{\it Introduction - }~Emerging microscopic phenomena feature astonishing collective properties, and trapped ion crystals offer unique opportunities for their 
classical or quantum simulation. This includes the study of defect formation in (structural) phase transitions \cite{islam2011onset, liprobing, ulm2013observation,pyka2013topological} but also magnetic phase transitions, which have been implemented first using a quantum simulator with two ions \cite{friedenauer2008simulating}, but now employ scaled-up systems for uncovering a  plethora of magnetic phases \cite{Blatt2012,zhang2017observation,zhang2017observation53,noel2022measurement,jurcevic2017direct,islam2011onset, kim2011quantum, schneider2012}. 
There is high interest in quantum simulations with even larger and eventually two-dimensional systems~\cite{ROOS,guo2023hundredions,alteeigenezickzackpaper}, as these would allow for analyzing frustrated spin models that feature elusive properties but are notoriously hard to predict from numerical simulations \cite{frustratedbermudez2011}. In a typical quantum simulation run, spins are initialized, their mutual interactions are switched on, and after a desired evolution time, the spin projection is recorded for each site by imaging laser-induced fluorescence on a spatial-resolving detector~\cite{bruzewicz2019trapped}. The interesting question is whether spin textures, as the outcome of a quantum simulation run, can be recorded alternatively. In particular, the question arises whether spin structures can be detected directly and in-situ by observing the interferences of photons coherently scattered by spins in a particular spin orientation.

Here, we demonstrate coherent scattering of light from an array of ions \cite{eichmann1993young,wolf2016visibility,cerchiari2021motion,obvsil2019multipath}, each of them acting spin-selectively as a single photon emitter (SPE). 
By measuring the intensity distribution  $I(\textbf{\textit{r}})$ in the far field
we are able to reveal 
the spin texture of the ion crystal using the benefits of Fourier optics. 
Investigating such interference patterns may offer advantages over conventional spatially resolved fluorescence detection, as this enables efficient far-field recording of a specific Fourier frequency without requiring knowledge of the individual spin orientations.
Moreover, we realize the in-situ detection \cite{hung2011extracting,putra2014optimally} in a background-free fashion, where the interference pattern is not perturbed by the exciting laser light.

We start with a short explanation of the interaction of the ion crystal with light and the emergence of an interference pattern $I(\textbf{\textit{r}})$, followed by the description of the experimental setup, including
the scheme for obtaining spin-dependent scattering by the trapped $^{40}$Ca$^+$ ions. We then analyze the fringe pattern —focusing on its contrast and the width and amplitude of its central maxima — to determine both, the SPE-array configuration and its spin order. Finally, we initialize the crystal with specific spin orientations and record their real-time evolution by observing the dynamics of the resulting interference patterns.  

{\it Interaction of the ion crystal with light - }~ We  describe the ion crystal by an effective model treating the atoms as localized photon sources that emit partially coherent light in a spin-selective manner. The intensity distribution $I(r)$ of an array of $N$ point-like particles is obtained by summing up their fields and taking the modulus square

\begin{equation}
\label{intensity n ions}
    I(\textbf{\textit{r}}) = c_1 \left| \sum\limits_{i=1}^{N} E_{i}(\textbf{\textit{r}})\right|^2 + c_2 \;
\end{equation}
Hereby, the term with prefactor $c_1$ denotes the properly normalized coherent part of the scattered light, whereas the prefactor $c_2$ represents the incoherently scattered contribution.  In the far-field, only the coherently scattered fields contribute to the grating-like fringe pattern, where the latter, displaying a contrast of $c_1/(c_1+c_2)$, encodes the spatial frequencies of the ion chain with particular spin orientation. 

\begin{figure}[ht!]
\centering
\includegraphics[width=0.95 \columnwidth]{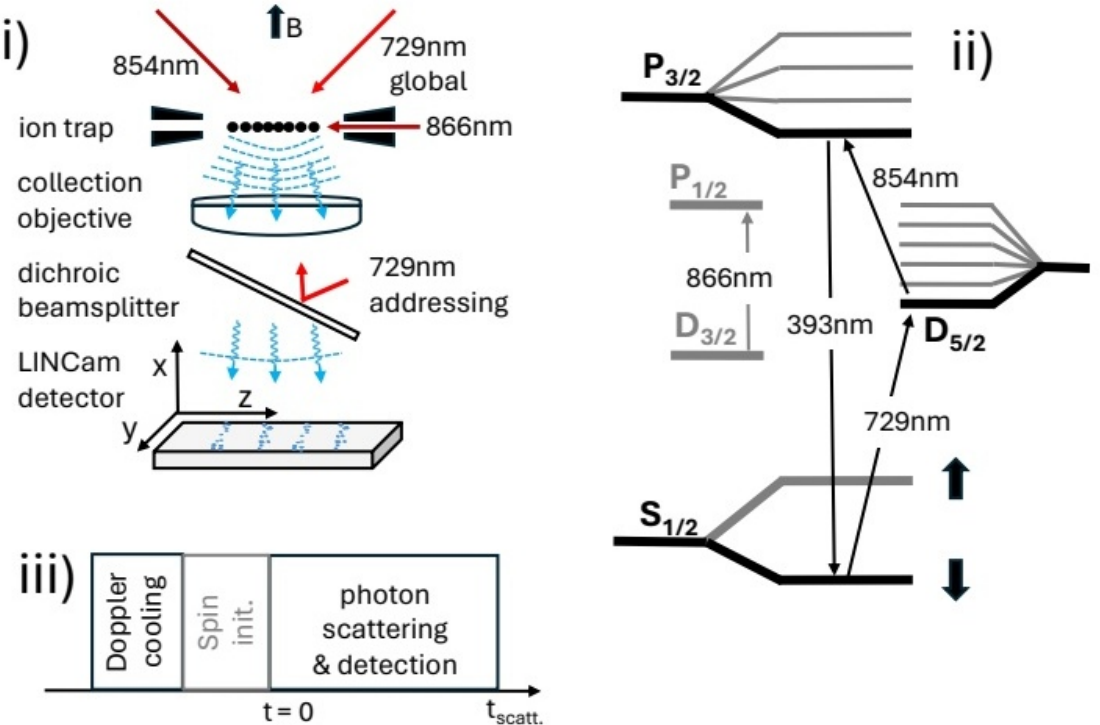}
\caption{i) Sketch of the experimental setup: A linear crystal of $^{40}$Ca$^+$ ions is exposed to laser fields (\SI{729}{nm}/\SI{854}{nm}/\SI{866}{nm}) with beam waist sizes of (\SI{25}{\upmu m}/\SI{80}{\upmu m}/\SI{80}{\upmu m}), respectively. Scattered photons near \SI{393}{nm} are collected by an NA$\,=\,$0.3 objective (Sill, S6ASS2241) and focused at a distance of \SI{127}{cm} on a slit of width \SI{1.4}{mm} (not shown in Fig. 1(i)) to filter out background photons. A multi-channel plate detector (LINCam from Photonscore Inc.) positioned \SI{170}{cm} downstream records the photons in the far field.
For single ion addressing, a beam near \SI{729}{nm} is reflected off a dichroic beam splitter and focused by the objective to a waist size of \SI{2}{\upmu m}.  ii) Levels and transitions in $^{40}$Ca$^+$; S$_{1/2}$ Zeeman levels m$\,=\,\pm$1/2 serve as spin up/spin down state, respectively. 
iii) Pulse sequence for spin-dependent detection: After Doppler cooling (\SI{397}{nm}/\SI{866}{nm}), spins are initialized by the addressing beam (\SI{729}{nm}); spin-dependent scattering is recorded under illumination by the global beams (\SI{729}{nm}/\SI{854}{nm}) while repumping residual population with \SI{866}{nm} from D$_{3/2}$ via the short-lived P$_{1/2}$ level to S$_{1/2}$.}
\label{fig:1}
\end{figure}

{\it Experimental setup - }~We confine linear crystals of $^{40}$Ca$^+$ ions in a linear Paul trap, built in X-blade geometry \cite{SebastianWolf_2019} with eleven DC segments. A radio frequency of $\Omega_{RF}/(2\pi)\,=\,\SI{30.04}{MHz}$ is applied to achieve secular frequencies along the two radial directions of $\omega_\mathrm{R1,R2}/(2\pi) = \{2.26,2.59\}$\,MHz. We adapt the voltages of the DC segments to trap linear crystals of three ions along the z-direction at an axial trap frequency of $\omega_\mathrm{z}/(2\pi) = \SI{0.72}{MHz}$ and relax to $\SI{0.58}{MHz}$ for trapping linear crystals with up to twelve ions. The two laser beams exciting the $^{40}$Ca$^+$ ions are confined to the x-z plane; hereby, the beam at \SI{729}{nm} exciting the quadrupole transition S$_{1/2} \rightarrow$ D$_{5/2}$ impinges the crystal under +\SI{45}{^\circ}, whereas the beam at \SI{854}{nm} driving the dipole transition D$_{5/2}\rightarrow$ P$_{3/2}$ illuminates the ions under \SI{-45}{^\circ}. 
The emitted photons near \SI{393}{nm} are collected in opposite x-direction by an objective of NA$\,=\,$0.3 and recorded in the far field by an ultra-fast spatial resolving camera (LINCam), see Fig.~\ref{fig:1}. 

During loading, we use the S$_{1/2} \rightarrow$ P$_{1/2}$ transition near \SI{397}{nm} for Doppler cooling, while a repumper near \SI{866}{nm} empties the metastable D$_{3/2}$ level, see Fig.~\ref{fig:1}(ii). A flip-mirror is placed in the setup to image the ion crystal on an EMCCD camera (both not shown in Fig.~\ref{fig:1}(i)), for probing the number of ions during loading prior and posterior to any of the measurements. A magnetic field of \SI{0.996}{mT} is applied in x-direction to provide a quantization axis and to separate the Zeeman levels of the S$_{1/2}$ state by \SI{27.92}{MHz}.

To realize spin-selective light scattering, the beam near \SI{397}{nm} is switched off, while the lasers at \SI{729}{nm} and \SI{854}{nm} are switched on to excite the narrow-band S$_{1/2}$, $m=-1/2$ ($\equiv \downarrow$) to D$_{5/2}$, $m=-5/2$ transition while simultaneously driving the transition near \SI{854}{nm}, which quenches the long-lived D$_{5/2}$ level via the P$_{3/2}$ state back to S$_{1/2}$, by emission of a single photon at \SI{393}{nm}, see Fig.~\ref{fig:1}(ii).  If the detuning, the Rabi frequencies and the polarization of the exciting laser beams are properly chosen, we implement an almost closed-cycle back into the S$_{1/2}$, $m=-1/2$ state. 
In this spin-selective scattering regime, the strengths of both laser fields are $\Omega_{729}/(2\pi) = 0.302 \pm 0.006\,\mathrm{MHz}$ and $\Omega_{854}/(2\pi) = 2.48 \pm 0.15\,\mathrm{MHz}$. They have been determined independently, by recording Rabi oscillations in the first case and observing the depletion dynamics of the D$_{5/2}$ level in the latter case. Under these conditions, we experimentally find a photon detection rate at \SI{393}{nm} of about 400~Hz per ion, 
free of background, with 20~Hz dark counts from the LINCam.

{\it Fringe contrast - }~ Only if the photons are scattered coherently, i.e., they have a fixed phase relation with respect to the incoming light field, 
the scattering process gives rise to interference fringes. We drive the 729 nm transition weakly and with a negative detuning $\Delta/(2\pi)\sim\SI{1.55}{MHz}$ 
to optimize both, the scattering rate and the fringe contrast. Modelling the three-level system as an effective two-level scheme \cite{marzoli1994laser,Reis1996} and using the above laser parameters, we find an effective linewidth  $\Gamma_{\text{theo}}/(2\pi)= 1.10 \pm 0.28\,\mathrm{MHz}$ overlapping with all of the ion crystals' vibrational red sidebands \cite{marzoli1994laser}. This leads to continuous sub-Doppler cooling \cite{janacek2015a} at a temperature of $80 \pm 5\,\mathrm{\upmu K}$, where the uncertainty stems from the uncertainties of the estimated Rabi frequencies. Experimentally, from resolved sideband spectroscopy, we find an ion temperature $\sim$~\SI{70}{\upmu K}, well below the Doppler limit of \SI{0.5}{mK}. Note that at this temperature, and because the trapping potential is highly anisotropic, the probability for swapping ions is negligible~\cite{abich2004}.  

In order to determine 
the fringe contrast
we recall that the intensity $I (\textbf{\textit{r}})$ coherently scattered by the ion array is given by~\cite{Itano1998,agarwal_quantum_2013}

\begin{align}
    I (\textbf{\textit{r}}) \sim & \sum\limits_{\substack{i, j=1\\
		i \neq j}}^{N}e^{-\mathrm{i}\textbf{\textit{q}}(\textbf{\textit{r}}_{i}-\textbf{\textit{r}}_{j})} \langle e^{\mathrm{i}\textbf{\textit{q}}(\textbf{\textit{u}}_j-\textbf{\textit{u}}_i)}\rangle \times C(N) \times \langle S_\mathrm{P,S}^{i\dagger}S_\mathrm{P,S}^j \rangle
		\label{Eq:2}
\end{align}

\noindent Here, the sum extends over $N$ ions at equilibrium positions $\textbf{r}_{i}$ with relative thermal fluctuations $\textbf{\textit{u}}_i$, $\textbf{\textit{q}} = \textbf{\textit{k}} - \textbf{\textit{k}}_L$ denotes the momentum transfer between the outgoing photon $\textbf{\textit{k}}=2 \pi \textbf{\textit{r}}/(\lambda \,r)$ and the two exciting photons $\textbf{\textit{k}}_L = \textbf{\textit{k}}_{729} + \textbf{\textit{k}}_{854}$ at \SI{729}{nm} and \SI{854}{nm}, respectively, and $S_\mathrm{P,S}^i$ ($S_\mathrm{P,S}^{i\dagger}$) represents the lowering (raising) operator of ion $i$ from the S$_{1/2}$ state to the P$_{3/2}$ state, mediating the effective two-level scheme.

On the right hand side of Eq.~(2), we identify  three terms which affect the contrast of the fringe pattern: 
The first term refers to the Debye-Waller factors DW$^{(ij)}=\langle e^{\mathrm{i}\textbf{\textit{q}}(\textbf{\textit{u}}_j-\textbf{\textit{u}}_i)}\rangle$
encoding the loss of coherence due to recoil-induced displacements. We have chosen a geometry which partially compensates for the photon recoil in the absorption and subsequent emission process ($\textbf{\textit{q}} \approx 0$, see Fig.~\ref{fig:1}(i)), giving rise to only small motional couplings characterized by the single ion Lamb-Dicke parameters $\eta_{(\mathrm{R1},\mathrm{R2},\mathrm{z})}$ = $\{$0.025, 0.023, 0.013$\}$. The DW$^{(ij)}$ factors are calculated by expressing thermal fluctuations in terms of eigenvectors of the ion crystal's motion~\cite{James1998}.
With increasing ion number, two effects enhance the DW$^{(ij)}$: Each vibrational mode is characterized by a larger crystal mass, thus less affected by photon recoil; and for N ions the remaining momentum transfer spreads over $3N-3$ vibrational modes. In addition, due to sub-Doppler cooling, the thermal fluctuations are small. The DW$^{(ij)}$ factors thus approach almost unity. 

\begin{figure}[ht!]
\centering
\includegraphics[width=1.15 \columnwidth]{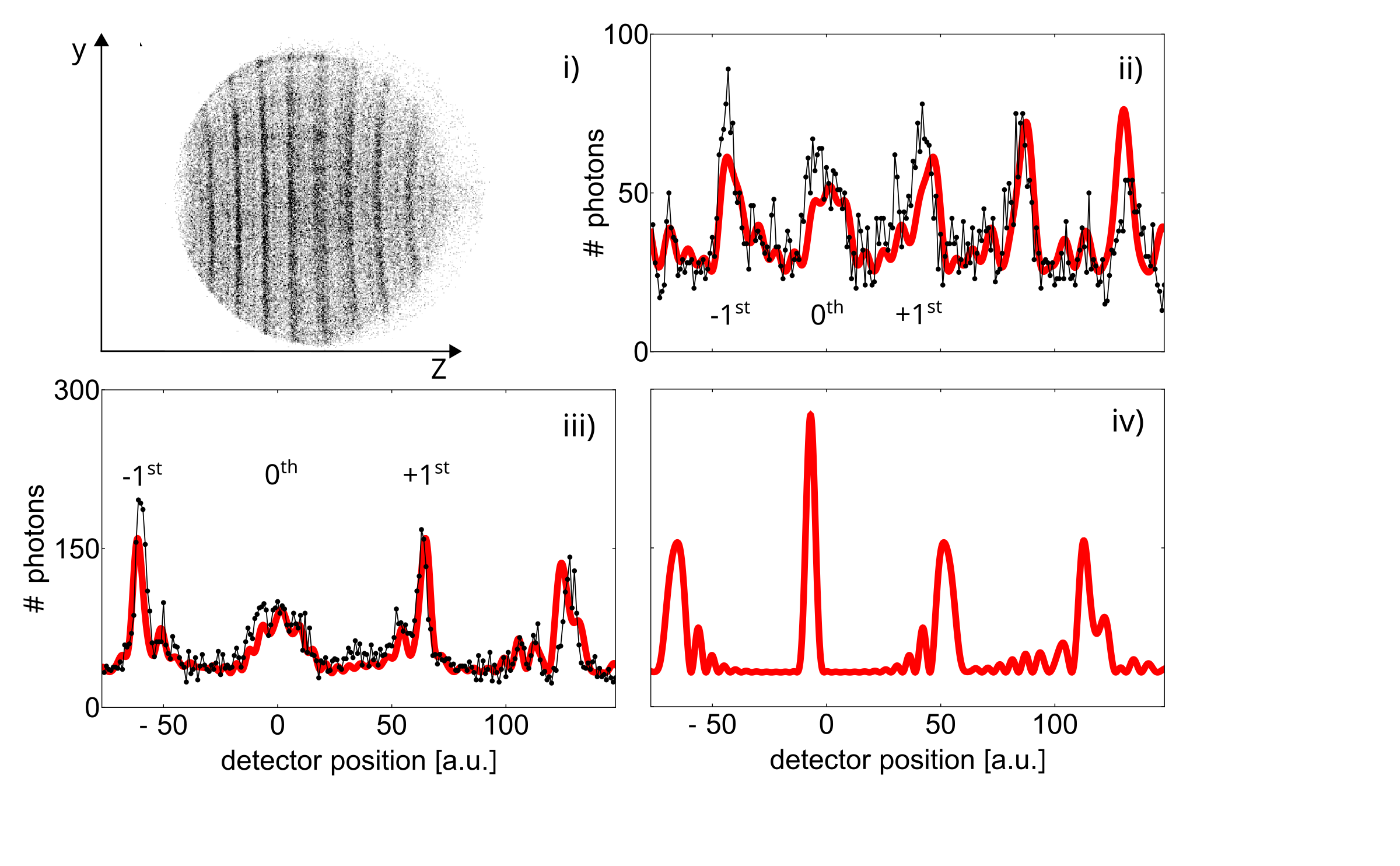}
\caption{(i) LINCam image (diameter 17~mm) with interference pattern from a $N=6$ ion linear crystal. Integrating over the y-axis leads to interference fringes displayed for linear crystals with (ii) $N=6$ and (iii) $N=12$ ions; data acquisition time is \SI{240}{s}. The model function (red) takes into account the independently determined ion positions, the laser beam profiles and a quadratic phase shift due to optical aberration, see text for details. (iv) Model function evaluated for the $N=12$ ion crystal, but excluding the quadratic phase shift.}    
\label{fig:2}
\end{figure}

The second term $C(N)$ considers the fraction of ion pairs in the spin $\downarrow$ state. As we do not initialize the spin states in the measurements shown in Fig.~\ref{fig:2}, each ion in the crystal may  either be in the $\downarrow$ state or in the $\uparrow$ state, where the latter scatters almost no light. Out of the $2^N-1$  spin configurations that scatter light, $N$ states with only one ion in the state $\downarrow$ do not contribute to the fringe pattern, leading to $C(N)=1-\frac{N}{2^N-1}$.

The last term depending on the coherences in the effective two-level scheme describes the amount of coherently scattered light by the ion chain, with $\langle S_\mathrm{P,S}^{i\dagger}S_\mathrm{P,S}^j \rangle = \langle S_\mathrm{P,S}^{i\dagger}\rangle \langle S_\mathrm{P,S}^j \rangle$ for $i \neq j$. 
The combination of all three terms in Eq.~\eqref{Eq:2} determines $c_1$ in Eq.~\eqref{intensity n ions}. We can also compute $c_2 = \sum \langle S_\mathrm{P,S}^{i\dagger}S_\mathrm{P,S}^i \rangle$, which represents the incoherently scattered light in Eq.~\eqref{intensity n ions}. Note, however, that in addition to the effects discussed above, we also have to take into account possible aberrations of the optical setup which affect $E_i(r)$ in Eq (1); this will be discussed in the next section. 



{\it Analysis of interference fringes -}~ We record interference patterns for linear crystals trapped in the same harmonic potential with four to twelve ions. At the detector we find interference patterns as shown in Fig.~\ref{fig:2}(i). Summing up the data in the central part of the fringe pattern along the y-direction for $N=6$ and $N=12$ ions, we obtain interference fringes as displayed in Fig.~\ref{fig:2}(ii) and (iii), respectively. (See online materials for the data set of $N=4$ up to $N=12$ ions \cite{verde2024}.) 
Here, the separation between the 0th diffraction order and the $\pm1^{st}$ diffraction order is determined by the inverse of the inter-ion separation $l_0$ which is \SI{3.26}{\um} at the center of a linear 12-ion crystal for an axial trap frequency of $\omega_\mathrm{z}/(2\pi)=$\SI{0.58}{MHz}. Yet, the inter-ion distance in a harmonic potential increases for ions further out of the center. Consequently, the $\pm 1^\mathrm{st}$ diffraction order is split up in several peaks, visible for instance in the case of a 12-ion crystal in Fig.~\ref{fig:2}(iii). Here, the $\pm 1^\mathrm{st}$ orders at positions $\pm$65 display a series of side peaks towards the center. 

Remarkably, also the 0$^\mathrm{th}$ diffraction order consist of many peaks instead of only a single narrow line. In an ideal optical setup, the 0$^\mathrm{th}$ diffraction order of a linear array of emitters placed along the z-axis would not display path length differences. We conjecture optical aberrations by the light collecting system, causing a parasitic phase shift for rays emitted by the ions displaced off the optical axis by a distance $z$. This aberration shows up especially when light is collected over a large field of view as is the case for crystals with large ion number. According to Zernike's description of lens imperfections, we introduce a quadratic parasitic phase shift $\phi(z_i)= a (z_i-z_0)^2$ for each ion, where $z_0$  accounts for an alignment offset. Moreover, we explicitly consider the spatial profile of the laser beam at \SI{729}{nm}, having a beam waist $w_0$, comparable to the length of the crystals. 

The electric field coherently emitted by an ion at $\textbf{\textit{r}}_i$ and recorded at $\textbf{\textit{r}}$ (see Eq.~\eqref{intensity n ions}) can therefore be written as:
\begin{equation}
\label{efield single ion}
    E_{i}(\textbf{\textit{r}}_i) \sim e^{\mathrm{i}a(z_i-z_0)^2} \times e^{-(\textbf{\textit{r}}_i/w_0)^2} \times e^{-\mathrm{i}\left(\frac{2\pi}{\lambda}\left|\textbf{\textit{r}}-\textbf{\textit{r}}_i\right|\right)},
\end{equation}
The first term describes the aberration-induced quadratic phase shift $\phi(z_i)$ and the second one the spatial profile of the \SI{729}{nm} laser, with the waist of the \SI{854}{nm} laser being much larger. Note that the wave front curvatures of the two exciting laser beams lead to a negligible phase shift. The last term denotes the path-length induced phase shift of a photon at $\lambda = \SI{393}{nm}$ emitted by an ion at $\textbf{r}_{i}$ and recorded at $\textbf{r}$. Optimum fit to the data is found for z$_0$ =~\SI{9}{\upmu m} and $a=\SI{0.015}{rad/\upmu m^2}$ (see Fig.~\ref{fig:2}(ii)), corresponding to a shift of $\pi$ for an ion off-center by $\pm\SI{15}{\upmu m}$. Taking the quadratic phase into account and adapting with three parameters for the overall amplitude, contrast and magnification of the interference fringes for the entire data set, we achieve very good agreement with the measured data, see Fig~\ref{fig:2}(ii),(iii). This model function is plotted in Fig~\ref{fig:2}(iv), now excluding the parasitic quadratic phases $\phi(z_i)$. Here, the 0$^\mathrm{th}$ diffraction order collapses into a narrow peak, whereas the $\pm$1$^\mathrm{st}$ orders display the inhomogeneous distribution of distances between ions, with a series of peaks in the direction towards the 0$^\mathrm{th}$ diffraction order. 

\begin{figure}[ht!]
\centering
\includegraphics[width=1.1 \columnwidth]{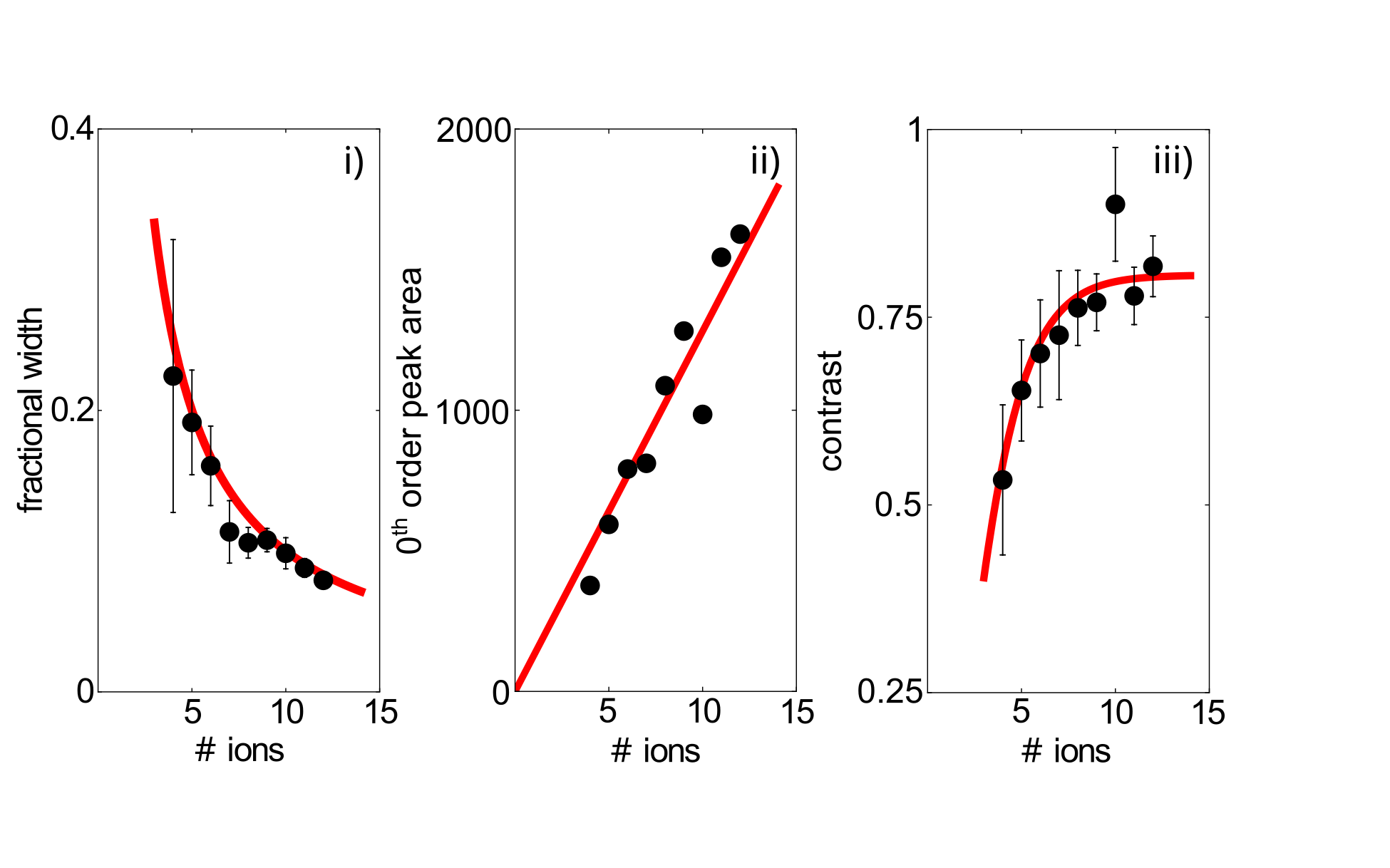}
\caption{(i) Fractional width of the $0^\mathrm{th}$ diffraction order, measured for crystals with 4 $\leq N \leq$ 12 ions trapped in a 0.58~MHz axial potential. The expected $1/N$ dependence is plotted (red), no free parameters. (ii) Photon counts integrated over the central peak showing a linear increase with $N$ (red). (iii) Contrast of the  interference fringes versus the number of ions $N$. The dependence on $N$ is described by a fit function, see text for details.}
\label{fig:3}
\end{figure}

{\it Scaling properties with ion number -}
We study the scaling of several quantities with the number of ions $N$ in Fig.~\ref{fig:3}. The fractional width of the $0^\mathrm{th}$ diffraction order shows the expected scaling with $1/N$, see Fig.~\ref{fig:3}(i); here, the fractional width is defined by the width of the main peak divided by the distance to the $\pm 1^\mathrm{th}$ order. Furthermore, we see from Fig.~\ref{fig:3}(ii) that the total photon count in the central peak increases linearly with $N$. Additionally, the contrast of the interference fringes increases with $N$ and saturates for $N>8$, see Fig.~\ref{fig:3}(iii). The high contrast stems from several reasons: As we do not initialize the spin states, the combinatorial factor $C$(N) approaches unity for $N \rightarrow \infty$; moreover, as outlined above, the Debye-Waller factors DW$^{(ij)}$ approaches unity with the experimental parameters of the setup.

{\it Observing spin textures -}~We extend our study to the regime of spin-selective coherent scattering. Now, the \SI{729}{nm} laser is tuned \cite{marzoli1994laser} sufficiently close to the S$_{1/2}$, $m = - 1/2$ to D$_{5/2}$, $m = - 5/2$ transition that only the $\downarrow$ state scatters. 
To record the real-time evolution of a $N=3$ spin system, we initialize the ion crystal's spin state and  observe the time-evolving fringe patterns arising from the different spin textures. Deterministic preparation is done by addressing single ion spins with an additional focused beam near \SI{729}{nm} (waist size $\sim \SI{2}{\upmu m}$), see Fig.~\ref{fig:1}(i). In this way, we are able to initialize all $2^N=8$ different spin-orientations.

\begin{figure*}[ht!]
\centering
\includegraphics[width=1.95 \columnwidth]{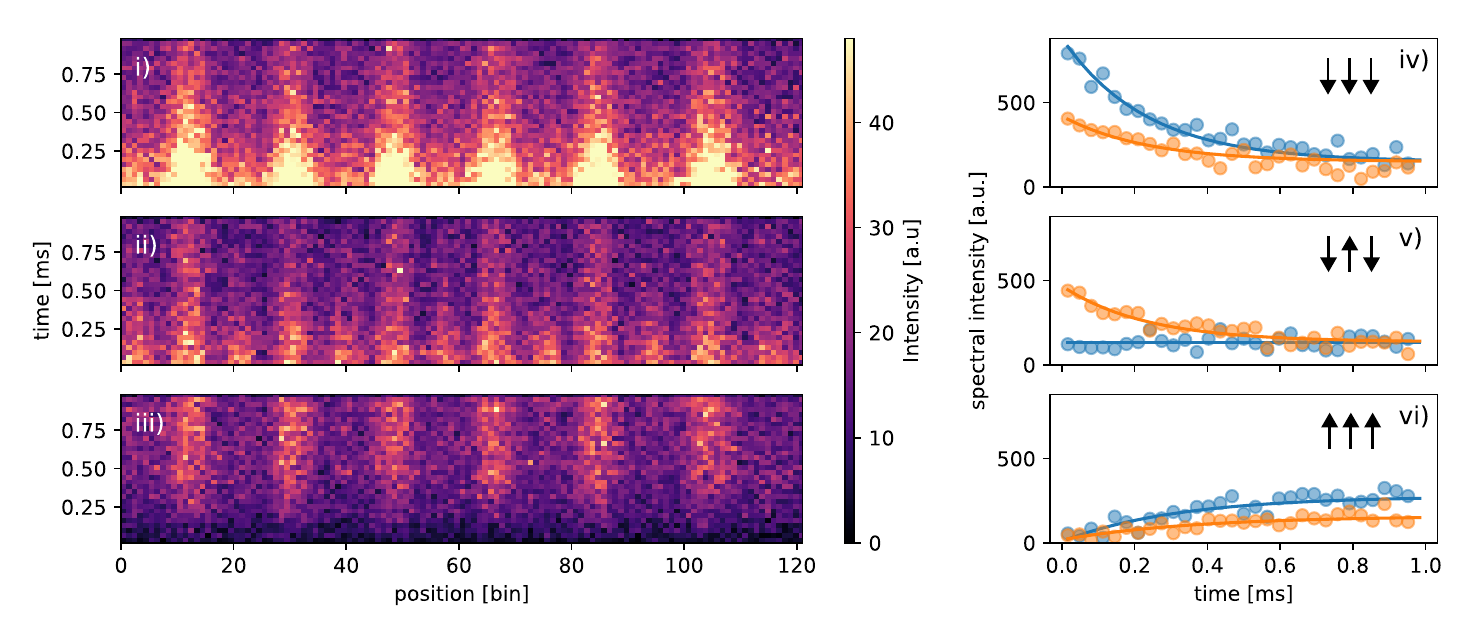}
\caption{Left side: Interference patterns observed from spin-initialized three-ion crystals with data acquisition time \SI{1}{ms}, repeated for 150,000 times; the three plots show the number of detected photons in each bin position (x-axis) vs. the scattering time t$_{scatt}$ (y-axis) for spin configuration initialization (i): $\{\downarrow,\downarrow,\downarrow\}$,  (ii): $\{\downarrow,\uparrow,\downarrow\}$, and (iii): $\{\uparrow,\uparrow,\uparrow\}$. Right side: The three plots display over time the intensity of the different spatial frequencies of the three-ion crystals, corresponding to different inter-ion distances with ions in state $\downarrow$; error bars are within the size of the dots. (iv): For the initial configuration $\{\downarrow,\downarrow,\downarrow\}$, we observe two spatial frequencies which decay as a function of time; (v): In the case of $\{\downarrow,\uparrow,\downarrow\}$, one Fourier frequency is present at the beginning and then decays, while the other one is barely visible throughout the observation time; (vi): Initializing the ion crystal to $\{\uparrow,\uparrow,\uparrow\}$, no spatial frequencies are observed initially, nonetheless they appear over time due a small leakage in the fluorescence cycle.}
\label{fig:4}
\end{figure*}

A time-resolved measurement on a spin-initialized crystal consists of repeating a sequence of three steps: Doppler-cooling, spin-initialization, and finally observation of the scattered photons at \SI{393}{nm} for \SI{1}{ms} while  continuously exciting the ions at \SI{729}{nm} and \SI{854}{nm}. As before, we integrate the photon detection events on the LINCam detector along the y-direction but now store in addition the time stamps of the recorded photons after the LINCam receives a trigger to start the photon detection. Such data allow for examining the time-evolution of the diffraction pattern what displays how the spin textures develop, see Fig.~\ref{fig:4}(i,ii,iii). This dynamics occurs due to a small leakage in the fluorescence cycle, pumping the spin textures eventually into equilibrium.

If we initialize the spins to the setting $ \{\downarrow,\downarrow,\downarrow\}$, all ions scatter light so that we observe two spatial Fourier frequencies in the fringe pattern, corresponding to the distance between the outermost ions and neighboring ions $2 l_0$ and $l_0=\SI{5.17}{\upmu m}$ at $\omega/(2\pi)=\SI{0.72}{MHz}$, respectively. They show up strongest directly after the initialization and decay for longer times, see Fig.~\ref{fig:4}(iv). As expected, the Fourier amplitude corresponding to a distance of $l_0$ is higher by a factor of $2.0 \pm 0.2$ compared to the one with distance $2 l_0$. On the other hand, if the crystal is initialized to $ \{\downarrow,\uparrow,\downarrow\} $, only one Fourier frequency appears corresponding to the distance $2l_0$. For an initialization to $ \{\uparrow,\uparrow,\uparrow\} $, no light is initially scattered. Yet, scattered light is observed with time due to the leakages in the fluorescence cycle. We prove that the experimental data display the expected Fourier components also for all other spin configurations, thus witnessing the initialized spin-texture (see online material \cite{verde2024}). 

When spin states are flipped due to optical pumping processes, we observe this (in situ) from the amplitude decay or the increase of a specific Fourier component; see Fig.~\ref{fig:4}(iv,v,vi). The rates for the dynamics $ \tau_{\downarrow \uparrow}  = 0.24 \pm 0.01\,\mathrm{ms}$ and in reverse $ \tau_{\uparrow \downarrow} = 0.33 \pm 0.01\,\mathrm{ms}$ are extracted from the entire data set. The recorded time variations of the respective Fourier amplitudes clearly show the ability to observe in situ the evolution of spin textures in an ion crystal. 

{\it Outlook -}~ We plan to employ collective spin-dependent scattering as an investigation tool for observing quantum phase transitions in analogue quantum simulations. 
Furthermore, we will improve the count rate using a pair of high-numerical aperture light collection objectives. Another challenge comes from the inhomogeneous distribution of the ion distances in a harmonic trap. Two solutions are intended: Either using a laser waist smaller than the total extension of the ion crystal such that only nearly equidistantly separated ions in the center part of the trap contribute to the interference fringes. Alternatively, we can tailor the axial potential in the micro-segmented trap ~\cite{wolf2016visibility,kaushal2020shuttling, ruster2014experimental} such that ion crystals with equidistant ions are formed. 

The source data used in this paper are available on RADAR \cite{verde2024} alongside corresponding supplementary plots.

We acknowledge financial support by the Deutsche Forschungsgemeinschaft (DFG, German Research Foundation), Project-ID 429529648 TRR 306 QuCoLiMa (Quantum Cooperativity of Light and Matter), by the Dynamics and Topology Centre funded by the State of Rhineland-Palatinate, the Erlangen Graduate School in Advanced Optical Technologies (SAOT) and thank Daniel Wessel and Dr. Jonas Vogel for careful reading. M. V., A. S. and B. Z. contributed equally to this work. 

\pagebreak

\end{document}